\documentstyle[psfig,prc,aps,rotate]{revtex}

\newcommand{\be}{\begin{equation}}
\newcommand{\ee}{\end{equation}}
\newcommand{\vs}{\vspace{7mm}}

\begin{document}

\begin{center}
{\bf \Large  Quark Coalescence and Hadronic Equilibrium}

\vs
{\sc T. S. Bir\'o\footnote{On leave from KFKI Budapest, Hungary}}

{\sc CERN Theory Division, CH-1211 Geneve 23, Switzerland}

\end{center}

\begin{small}
{\bf Abstract}
In a simplified model we study how close can the result of
a fast non-equilibrium process, like hadronization of
quark matter in an expanding fireball, come to  hadronic composition
in equilibrium.  In a chemical approach of
in-medium quark fusion \hbox{($A + B \rightarrow C$)}
we find that this depends on 
the time-integrated rate per volume multiplied by the conserved
number of constituents. Ideal hadron gas equilibrium
constrained by conservation laws, the fugacity parametrization,
as well as linear and non-linear coalescence approaches
are recognized as different approximations to this chemistry.
It is shown that color confinement requires a dependence of the hadronization
cross section on quark density.
\end{small}

\vs

\vs
Since strangeness enhancement had been suggested as a possible signature
of quark-gluon plasma formation in heavy-ion collisions, there is
a particular attention to the question of hadronic and quark level
equilibrium in these reactions 
\cite{Rafelski:1982pu,Biro:1982zi,Biro:1983gh,Koch:1986ud}. 
Recent claims of describing data
at AGS as well as at CERN SPS by constrained hadronic equilibrium
(i.e. using a common volume, temperature and only chemical
potentials of conserved quantities for the computation of hadron
numbers) ``quite'' satisfactorily 
\cite{Braun-Munzinger:1999qy,Becattini:1998we,Becattini:1998ii,Cleymans:1999yf,Davidson:1992pa} 
led to some astounishment
in the community: the total reaction time at SPS energy is namely
too short for achieving hadronic equilibrium by hadronic conversion
processes alone.
Pushing aside the faux pais, that this ``satisfactory'' equilibrium
description agrees with experimental data for a few ratios only within a
factor of two while the general agreement is on the 30\% level, 
one faces the question, how can this
simple idea work even so well. The time argument above would exclude this.

\vs
Meanwhile there is a number of non-equilibrium approaches
to this problem ranging from the use of phenomenological
parameters describing the deviation on the hadronic level
(fugacities) 
\cite{Rafelski:1996zj,Letessier:1999sz}
through linear \cite{Bialas:1998ea} and nonlinear 
(constrained) coalescence models
\cite{Biro:1995mp,Zimanyi:1997zz},
and mixed quark-hadron chemistry \cite{Biro:1999cx} 
up to very detailed microscopic simulations of the hadronization
(parton cascades \cite{Geiger:1992nm,Geiger:1993cm,Geiger:1997ym,Geiger:1998fq}
VENUS \cite{Werner:1987ze},
HIJING \cite{Eskola:1996bp,Wang:2000uj},
RQMD \cite{Sorge:1995vv,Sollfrank:1999zg},
UrQMD \cite{Antinori:1999rw,Bass:1996ud,Bass:1999vz,Bleicher:1999cw}, 
string model
\cite{Amelin:1990vp,Dean:1992vj,Bass:1998ca,Csizmadia:1999vp},
chromodielectric model \cite{Traxler:1999bk}).
The basic physical concepts of these approaches oppose hadronic
equilibrium; it might be reached in course of time, but for some 
components it is actually
not. For recent transport simulations at SPS 
energies see \cite{Bratkovskaya:2000qy}.
There are also recent doubts whether chemical equilibrium can be reached at
RHIC \cite{Lindenbaum:2000uq}.

\vs
The purpose of this letter is to identify a useful quantitative
factor in describing the approach to hadronic equilibrium
states with quark level processes instead of hadronic ones in a fast
expanding fireball. In order to concentrate on this quantifying
effort of the time evolution of a complicated system we shall
deal with a simple fusion process on the constituent level: 
$$ A + B \longleftrightarrow C.$$
An example is given by the fusion of an $s$ quark and a $ud$
diquark to $\Lambda$ particle. In this case baryon charge and
strangeness conservation constrains the equilibrium.

\vs
In such a process the energy-impulse balance is assumed to be satisfied by the
surrounding medium, we are interested in the time evolution of
the number of final state products $N_C(t)$. The equilibrium number
of precursors $N_{A,eq}$ and $N_{B,eq}$ - in a quark matter scenario - will 
eventually be taken as zero. The initial state is assumed to be far from this
equilibrium, it is oversaturated by $A$ and $B$. The simplest is
to consider no $C$ particles at all in the beginning.
As the reaction proceeds there will be $C$ type particles
at the end (or equilbrium  will never be reached), they are hadrons.
The cross section for such a reaction - also as a simplification -
can be taken as a constant parameter, but temperature and reaction volume
must be considered as time dependent quantities.
We shall point out, however, that the assumption of constant
cross sections actually contradicts to the confinement principle
in quark matter scenarios, so in that case the hadronization
cross section has to be quark density dependent. This extra dependency,
not derivable from perturbative or lattice QCD so far, will be referred
as ``dynamical confinement'' in this paper.

\vs
In the followings -- after a brief presentation of general
chemical equations -- we shall consider the simple model of
hadronization due to coalescence (fusion in the presence of a medium). 
Comparison of different limiting cases 
(allowing or excluding the precursors $A$ and $B$ in equilibrium) and
estimates of the characteristic deviation from  hadronic
equilibrium with and without dynamical confinement will close this study.


\vs
In the chemical approach to this process all particle numbers are time dependent.
The entropy grows during a spontaneous approach towards equilibrium. 
A general reaction can be characterized by stochiometric 
coefficients $\alpha_i^{(r)}$ in a fashion arranged to zero:
$$ \sum_i \alpha_i^{(r)} N_i \longleftrightarrow 0$$
for each reaction $r$.  In this arrangement the $\alpha_i^{(r)}$-s 
can be negative as well as positive.
An important quantity with respect to approaching chemical equilibrium
is the activity of a reaction,
\be x^{(r)} = \sum_i \alpha_i^{(r)} \frac{\mu_i}{T}.\ee 
It describes the (un)balance of gain and loss terms.
Rate equations summarize contributions from each reaction channel,
\be 
 \dot{N}_i = \sum_{(r)} \alpha_i^{(r)} R^{(r)} 
 \left( 1 - e^{x^{(r)}} \right). 
 \label{RATE_EQ}
\ee
Here the factors $R^{(r)}$ depend on numbers of other particles,
cross sections and volume. Due to known (e.g. adiabatic) expansion
a relation between volume $V$ and temperature $T$ is given.

\vs
In simple scenarios of expansion $VT^{3/2} = $ const. (massive ideal gas
expands adiabatically, no back effect from chemistry is taken into
account). Now
\be N_i = N_{i,eq} e^{\mu_i/T} \ee
and 
\be 
   N_{i,eq} = V d_i \left( \frac{m_iT}{2\pi}  \right)^{3/2} e^{-m_i/T}. 
 \label{BOLTZMANN}
\ee
In simple scenarios like this the time dependence of particle numbers, 
$N_i(t),$ is entirely due to that of the \hbox{$(\mu_i(t)-m_i)/T(t)$.}
The equilibrium ratio,
\be
R_{{\rm eq}} = \frac{N_{A,{\rm eq}} N_{B,{\rm eq}}}{N_{C,{\rm eq}}}
= \frac{d_Ad_B}{d_C} \left( \frac{m_Am_B}{m_C} \right)^{3/2}
VT^{3/2} e^{(m_C-m_A-m_B)/T}
\label{EQ-RATIO}
\ee
is nevertheless constant for adiabatic expansion ($VT^{3/2}$=const.)
and composite final products ($m_C=m_A+m_B$).

\vs
All rate equations drive the system towards chemical equilibrium.
The total entropy production can be obtained from
\be TdS + \mu_i dN_i = dE + pdV = 0 \ee
for adiabatic flow of an ideal (``dry'') fluid. One obtains
\be \dot{S} = - \sum_i \frac{\mu_i}{T} \dot{N}_i =
\sum_r R^{(r)} x^{(r)} \left( e^{x^{(r)}} - 1 \right) \ge 0. \ee
In chemical equilibrium $ x^{(r)} = 0 $ for each reaction, i.e.
\be \sum_i \alpha_i^{(r)} \mu_i = 0. \ee
There are as many linear constraints for the chemical potentials $\mu_i$ 
as many reactions.
If this is less than the number of particle sorts,
the equilibrium state is undetermined.
If this is greater than the number of particle sorts,
the reactions cannot be all independent: there are conserved
quantities (e.g. from $\dot{N}_1 = \dot{N}_2$ it follows that
$B = N_1 - N_2$ is conserved).
In this latter case all $\mu_i$ can be expressed by a few numbers
corresponding to these conserved quantities. This is the constrained
equilibrium.

\vs
Let us denote the conserved quantities by $Q_a$ (a = baryon, strangeness, etc.).
The corresponding charge of the i-th particle sort let be $q_{i,a}$.
We have
\be \sum_i q_{i,a} \dot{N}_i = 0 \ee
Substituting the rate equations (\ref{RATE_EQ}) we get
\be \sum_i \sum_r \alpha_i^{(r)}q_{i,a} R^{(r)} \left(1-e^{x^{(r)}} \right)
= 0 \ee
Since conservation is fulfilled even out of equilibrium, we conclude that
\be 
  \sum_i \alpha_i^{(r) } q_{i,a} = Q^{(r)}_a = 0, 
  \label{CONSERV}
\ee
i.e. there is no charge unbalance in any of the reactions.
This fact allows us to redefine the chemical potentials as
\be \mu_i = \sum_a q_{i,a} \mu_a + \tilde{\mu}_i.\ee 
The first contribution, depending on only as many ``chemical potentials''
as conserved charges we have, does not change the activities:
\be x^{(r)} = \sum_i \sum_a \alpha_i^{(r)} \frac{q_{i,a}\mu_a}{T}
+ \sum_i \alpha_i^{(r)} \frac{\tilde{\mu}_i}{T}. \ee
Since the first part is zero due to eq.(\ref{CONSERV}), 
chemical equilibrium (vanishing activities) means
\be \sum_i \alpha_i^{(r)} \frac{\tilde{\mu}_i}{T} = 0. \ee
For a well-determined or overdetermined system the only equilibrium
solution is given by $\tilde{\mu}_i=0.$
In this (constrained) equilibrium state the original chemical potentials
are a linear combination of a few $\mu_a$-s, associated to conserved charges,
\be \mu_{i,eq} = \sum_a q_{i,a} \mu_a. \ee 

\vs
Whether this chemical equilibrium state can be reached in a finite time
depends on
the rates $R^{(r)}$ and activities $x^{(r)}$. 
In this respect whether a quantity is conserved or not is determined by
the set of rate equations. This way it happens that not only real conserved
charges but also the number of each constituent quark may be conserved
in a sudden hadronization process, like ALCOR assumes this.

\newpage
For ideal massive particles
(Boltzmann distribution)
\be \mu_i = T \log \frac{N_i}{N_{i,eq}}. \ee 
The activities hence contain ratios of equilibrium numbers in each
reaction (mass action law, detailed balance). For a simplified
expansion picture these ratios are (nearly) constant. The factors $R^{(r)}$
contain precursor numbers, averaged cross sections (fluxes) and
the reaction volume. In an expanding system its time-dependence is 
far from trivial.

\vs
For a general set of reactions and particle sorts the time evolution
can only be followed numerically. In order to understand, however,
possible deviations from equilibrium in a final state - even after infinitely
long time - we consider an oversimplified system in the followings.


\vs
In the case of only one reaction channel,
$ A + B \longleftrightarrow C,$ we denote the equilibrium ratio by
$R_{{\rm eq}}$ (cf. eq.(\ref{EQ-RATIO})).
Since $\alpha_A=\alpha_B=1$ and $\alpha_C=-1$ for this very reaction,
we have $\dot{N}_A=\dot{N}_B=-\dot{N}_C$. From this two
conservation laws follow,
\be N_A + N_C = N_A(0) = N_0, \label{CONA} \ee
\be N_B + N_C = N_B(0) = N_0. \label{CONB} \ee
For considering an evolution towards equilibrium the
specific rate,
\be \nu =  \frac{\langle \sigma v \rangle}{V}, \ee
is a useful quantity. It occurs in the
rate equation 
\be 
   \dot{N}_C = \nu \left( N_A N_B - R_{eq} N_C  \right). 
\ee
Substituting the conservation laws (\ref{CONA},\ref{CONB}) we arrive at
\be 
  \dot{N}_C = \nu \left((N_0-N_C)(N_0-N_C)- R_{eq} N_C \right)
  = \nu (N_C-N_+)(N_C-N_-). 
  \label{SIMPLE_RATE_EQ}
\ee
This rate is vanishing for the particle numbers $N_C=N_{\pm}$ with
\be N_{\pm} = R_{eq}/2 + N_0 \pm \sqrt{R_{eq}^2/4 + R_{eq}N_0}. \ee
Here $N_-$ is a stable, $N_+$ is an unstable equilibrium point.
For a non-relativistic, adiabatic expansion
\be VT^{3/2} = {\rm const.} \ee
and hence  $R_{eq} = const.$ for constituent hadrons 
(those with mass $m_C = m_A + m_B$).
A realistic limit is $R_{eq}=0$ when no $A$ or $B$ particles are
allowed in equilibrium. In particular this is the case for
quark confinement.

\vs
The simple rate equation (\ref{SIMPLE_RATE_EQ}) can be solved
analytically giving
\be  
   \int_{N_C(0)}^{N_C(\infty)} \frac{dN_C}{(N_C-N_+)(N_C-N_-)} =
   \int_{t_0}^{t_{br}} \nu dt = r. 
\ee
Here $t_0$ is the initial time, when the fireball contains no hadrons,
$N_C(0)=0$. $t_{br}$ is the break-up time, beyond which no chemical
reactions (even no particle collisions) occur. In some cases it is
large compared to characteristic times and can be approximated by
infinity.
Now using the identity
\be 
   \frac{1}{(N_C-N_+)(N_C-N_-)} = \frac{1}{N_+ - N_-} 
   \left( \frac{1}{N_C-N_+} - \frac{1}{N_C-N_-} \right) 
\ee
we arrive at
\be 
  \left.  \log(N_C-N_+) - \log(N_C-N_-) \right|_0^{N_C(\infty)} =
          (N_+-N_-) r . 
\ee
This leads to the solution
\be 
   N_C(\infty)= N_+N_- \frac{1-K}{{N_+} - K {N_-}} 
\ee
with
\be K =  \exp \left( -  (N_+-N_-)r \right). \ee
Here $N_-$ is the constrained equilibrium value, achieved for $K=0$.
There are two alternative forms of this relation 
derived by using $N_+N_-=N_0^2$.
The first one is the coalescence form,
\be N_C(\infty) = N_0^2 \frac{1-K}{N_+-KN_-} = C(K,N_0) N_0^2 =
C(K,N_0) N_A(0) N_B(0),  \ee
the second one is the equilibrium form,
\be N_C(\infty) = N_- \frac{1-K}{1-K(N_-/N_0)^2} = \gamma_C N_-
= \gamma_C N_{C,{\rm eq}}. \ee
In particular $ N_C(\infty) = N_-$ in the $K=0$ (infinite rate) case.
On the other hand in the case of confinement $R_{eq}=0$, from which
$N_+=N_-=N_0$ follows. 

\vs
This agrees with the ALCOR assumption: after long enough time
all quarks A and B are converted into hadrons C. This means $N_- = N_0$,
satisfied for $R_{eq}=0$, i.e. when the quark equilibrium numbers
are zero (confinement in equilibrium). Then ALCOR ends up with the
equilibrium number. (Note: ALCOR does not take into account the gain/loss
balance as if it were infinite activity in favor of hadrons - against quarks.
This feature is equivalent to putting $R_{eq}=0$.)


\vs
Without quark confinement, but $(N_+-N_-)$ 
small (${\cal O}(R_{eq})$), i.e. when precursors 
are suppressed in the equilibrium state,  
we consider a negligible, but not zero $R_{{\rm eq}}$.
In the limit of $R_{{\rm eq}} \rightarrow 0$ the precursors
$A$ and $B$ are suppressed in equilibrium, according to the
static confinement principle. This is, however, insufficient
for hadronizing all quarks in an expanding matter.
We show this in the followings.

\vs
We expand $K$ in terms of $(N_+-N_-)$ which is in the same order of
magnitude. We get
\be K = \exp(- (N_+-N_-)r) = 1 - r(N_+-N_-) \ee
and the final hadron number becomes
\be 
  N_C(\infty) = N_0^2 \frac{r}{1+rN_0}.
  \label{BASIC}
\ee

\vs
Eq. (\ref{BASIC}) -- valid if the precursors are suppressed in
chemical equilibrium (and this is all we know from lattice QCD) --
has different readings. For $r$ small it comes close to the linear 
coalescence equation
\be
  N_C(\infty) = r N_0^2 = r N_A(0) N_B(0).
  \label{LINEAR}
\ee
(Note that $N_A(0)=N_B(0)=N_0$.)
ALCOR uses extra modification factors $b_A$ and
$b_B$ by assuming the modified coalescence rate
\be
  N_C(\infty) = r b_A b_B N_0^2.
  \label{NON_LINEAR}
\ee
To start with it seems to be rather similar to the coalescence picture.
Due to conservation, however, ALCOR obtains
\be
  N_C(\infty) = r b_A b_B N_0^2 = \frac{r}{1+rN_0}N_0^2
  \label{ALCOR}
\ee
and determines the factors $b_i$ accordingly (depending on $N_0$).
It leads exactly to the constrained equilibrium result in the
limit $rN_0 \rightarrow \infty$:
\be
  N_C(\infty) =  N_0.
  \label{EQUIL}
\ee
So, reversing the argument, we conclude that for quark matter hadronization
$r$ has to diverge in order to reflect confinement in the end stage.

\vs
Finally the chemistry approach gives the interpretation of
the fugacity due to
\be
  N_C(\infty) = N_0^2 \frac{r}{1+rN_0} = \gamma_C N_0.
  \label{FUGA}
\ee
In fact the difference between these approaches is more pronounced
by inspecting chemical potentials instead of final numbers.
In the discussed simple case the chemical potential of the
(final state) hadron is given by
\be
 \frac{\mu_C}{T} = \log \frac{N_C(\infty)}{N_{C,eq}}.
 \label{BOLTZMANN_CHEM_POT}
\ee
Substituting our result for $N_C(\infty)$ it splits to contributions
with different physical meanings:
\be
 \frac{\mu_C}{T} = \log \frac{N_-}{N_{C,eq}} \, + \, 
  \log \frac{N_C(\infty)}{N_-}
 = \frac{q_{C,1}\mu_1 + q_{C,2}\mu_2}{T} + \log \gamma_C.
 \label{CHEM_POT_C}
\ee
The conserved quantities (charges) carried by the hadron $C$ 
are those involved in the identities respected by the rate equations,
$N_A(t)+N_C(t)=N_A(0)=N_0$ and $N_B(t)+N_C(t)=N_B(0)=N_0$. In this case 
$q_{C,1}=1$ and $q_{C,2}=1$, respectively. The fugacity parameter
(as used by Rafelski et al.) is the remaining part of the formula,
in equilibrium it is $\gamma_C=1$.


\vs
Finally we note that the ALCOR approach (constrained coalescence)
leads to equilibrium result only if a single reaction channel
is considered, like in the case discussed in the present paper.
For several, competing channels the relative hadron ratios
are different for ALCOR and constrained equilibrium.
This best can be seen by inspecting the corresponding $\mu_i$
chemical potentials to the pre resonance decay hadron numbers
as a function of strangeness in the ALCOR simulation.
Slight deviations from the
perfect straight lines, which would mark the constrained equilibrium
in the ideal case, can be observed \cite{Biro:S2000}.

\vs
In case of confinement for particles $A$ and $B$ 
the equilibrium ratio $R_{eq}$ vanishes and
$N_-$ has to be replaced by $N_0$ in the above formula.
Exactly this way are usually the chemical potentials associated to
conserved charges, $\mu_1$ and $\mu_2$, determined.
The fugacity factor 
\be
  \gamma_C = \frac{1-K}{1-K(N_-/N_0)^2},
\ee
in only static confining scenarios reduces to
\be
  \gamma_C = \frac{rN_0}{1+rN_0} < 1.
\ee
This must be viewed as an indication for the necessity of considering
dynamic confinement in hadronization.

\vs
There is no difference between ALCOR and constrained equilibrium
in this simple model if confinement is implemented: 
both lead to $\gamma_C=1$ eventually. The linear coalescence
picture on the other hand assumes $rN_0 \ll 1,$ corresponding to a
rather impressive deviation from equilibrium. 
Therefore it is interesting to estimate the
factor $rN_0$ for different heavy-ion experiments.

\vs
We continue by estimating the integrated specific rate $r$.
Its value depends not only on cross section(s) -- first assumed
to be constant and later density dependent -- but also on the
expansion and cooling of the reaction zone. We consider here two
limiting cases: that of the one dimensional expansion with a
rather relativistic equation of state (ideal gas with 
particles having negligible masses) and that of three dimensional
expansion of a massive, non-relativistic ideal Boltzmann gas. The first
case is thought to be rather characteristic for RHIC and LHC, the
latter rather for $SPS$ and $AGS$ energies.

\vs
During a one-dimensional expansion the volume grows linearly in time,
\hbox{$ V = \pi R_0^2 t$,}
and the (thermal) average of relative velocities is near to the
light speed $\langle v \rangle = 1$. In this case we obtain
\be
 r = \int_{t_0}^{t_{br}} \frac{\sigma}{\pi R_0^2} \frac{dt}{t}
 = \frac{\langle \sigma v_0 \rangle}{V_0} t_0\log \frac{t_{br}}{t_0}.
 \label{RHIC_rate}
\ee
The time-integrated specific rate grows with the break-up time $t_{br}$
monotonically, it can eventually be infinite, if there is no thermal break-up
of the expanding system. In that case $\gamma_C=1$, equilibrium
is achieved.
For a three-dimensional expansion one gets $V=V_0(t/t_0)^3$. Furthermore
a massive, non-relativistic ideal gas satisfies $VT^{3/2}=V_0T_0^{3/2}$
during an adiabatic expansion. The average relative velocity is 
temperature dependent like 
\hbox{$\langle v \rangle \sim \sqrt{T/m} \sim 1/t$,}
with some reduced mass of the fusing pair $m$. This together leads to
the estimate
\be
  r = \frac{\langle \sigma v_0 \rangle}{V_0} \int_{t_0}^{t_{br}}
t_0^4 \frac{dt}{t^4}.
\ee
Evaluating the integral we arrive at
\be
 r = \frac{1}{3} \frac{\langle \sigma v_0 \rangle}{V_0} t_0 \left( 
 1 - \left( \frac{t_0}{t_{br}}\right)^3 \right).
 \label{SPS_rate}
\ee
It is interesting to note that $r$ is finite even without break-up,
i.e. in the $t_{br} \rightarrow \infty$ limit.
This also means that the non-relativistic, 3-dimensional spherical
expansion scenario with constant hadronization cross section cannot
comply with quark confinement. In this case $\sigma$ has to be
time dependent at least as $\sigma \sim t^3 \sim V \sim 1/n$,
i.e. the hadronization cross section has to be inversely proportional
to the density of quarks. Exactly this assumption was made in
transchemistry \cite{Biro:1999cx}.

\vs
In orther to simplify the further discussion
 we relate the thermal average of the rate at the beginning
of the expansion to estimates of the collision time due to mean free path,
\be
  t_{c} = \frac{1}{\langle \sigma v_0 \rangle \rho_0},
  \label{ORIG_COLL_TIME}
\ee
with density $\rho_0 = 2N_0/V_0$. This leads to
\be
  \frac{\langle \sigma v_0 \rangle}{V_0} = \frac{1}{2N_0t_c}.
   \label{INIT_RATE}
\ee
The quantity describing the deviation from constrained equilibrium
for hadrons then is given as follows:
\be
 rN_0 = \frac{t_0}{2t_c} \log \frac{t_{br}}{t_0}
 \label{DEV_RHIC}
\ee
for one-dimensional relativistic flow and
\be
 rN_0 = \frac{t_0}{6t_c} \left(
 1 - \left(\frac{t_0}{t_{br}}\right)^3 \right)
 \label{DEV_SPS}
\ee
for three-dimesional non-relativistic flow of massive constituents.
The general dependence of the fugacity on the estimated collision
time is in both cases
\be
 \gamma_C = \frac{1}{1+\beta (t_c/t_0)},
 \label{LAMBDA}
\ee
just the constant $\beta$ differs. For the realtivistic, one-dimensional
flow case it is
\be
 \beta = \frac{2}{\log \frac{t_{br}}{t_0} }
 \label{RHIC_BETA}
\ee
and for three-dimensional spherical flow of massive particles it is
given by
\be
  \beta = \frac{6}{1 - \left( \frac{t_0}{t_{br}} \right)^3 }.
  \label{SPS_BETA}
\ee
From these formuli one realizes that for $t_{br}=t_0$,
i.e. in case of immediate break-up the chemical equilibrium cannot
be approached, no hadron $C$ will be formed and $\gamma_C$ remains
zero. The opposite limit, $t_{br} \rightarrow \infty$ is less
uniform: while in the relativistic one-dimensional scenario 
equilibrium is always
reached if enough time is given ($\gamma_C(\infty)=1$), this is not
so for a stronger, more spherical expansion. 
In the latter case $\gamma_C$ tends to
a finite value,
\be
 \gamma_C(\infty) = \frac{1}{1+6t_c/t_0},
\label{FINITE-GAMMA}
\ee
which is in general less than one. It means that chemical equilibrium
cannot be really reached even in infinite time. (An analysis of 
proton -- deuteron mixture with a similar conclusion was given in
Ref.\cite{Biro:1984yz}.) For early break-up, $t_{{\rm br}}-t_0 \ll t_0$,
both scenarios lead to
\be \beta = \frac{2t_0}{t_{{\rm br}}-t_0}. \ee


\vs
In quark matter hadronization scenarios eq.(\ref{FINITE-GAMMA})
contradicts to the quark confinement principle. In order to
reach $\gamma_C(\infty)=1$ a dynamical confinement mechanism,
causing the quark fusion cross section in medium to be
(at least) inversely proportional to the quark density, has to
be taken into account besides the equilibrium suppression of
quark number (referred in this paper as ``static confinement'').

\vs
Let us assume that $A$ and $B$ are quark level (colored) objects,
and the fusion cross section scales as
\be
\sigma = \sigma_0 \left( \frac{V}{V_0} \frac{2N_0}{N_A+N_B} \right)^{1+\epsilon}
\label{SIGMA-SCALES}
\ee
with some extra power $\epsilon$. The solution of the rate equation
in the $R_{{\rm eq}}=0$ case (both static and dynamic confinement) becomes
\be
N_C = N_0 \left( 1 - (1-\delta)^{1/\epsilon} \right),
\ee
with
\be
\delta = \frac{t_0}{6t_c} \left( \left( \frac{t}{t_0} \right)^{3\epsilon}
- 1 \right).
\ee
For $\epsilon = -1$ (without dynamical confinement) one gets back
the result discussed so far. In the limiting case $\epsilon=0$
(minimal dynamical confinement) one obtains
\be
N_C = N_0 \left( 1 - \left( \frac{t_0}{t} \right)^{\frac{t_0}{2t_c}} \right).
\label{TRANS-CHEM}
\ee


\vs 
Fig.1 plots the dependence of the non-equilibrium fugacity 
factor $\gamma_C$ on the scaled break-up
time $t_{br}/t_0$ for these scenarii: one and three dimensional scaling
expansion, in the three-dimensional case both for static and 
dynamic confinement (color density dependent
quark fusion cross section).  The uppermost curve, showing the fastest
equilibration belongs to a 3-dimensional expansion scenario with a
cross section proportional to the inverse of the quark density
(cf. eq.(\ref{TRANS-CHEM})). The lowest curve, saturating well below
the equilibrium value, corresponds to the same expansion with
static confinement only. The middle curve shows the static RHIC
scenario, reaching equilibrium (vanishing quark number) eventually,
but quite slowly. For all calculations we used the ratio
$t_0/t_c=3$ for the sake of simplicity. (The smaller $t_0$, the higher
density, therefore the shorter inter-collision time $t_c$
initially.)


\begin{figure}
\centerline{\rotate[r]{\psfig{figure=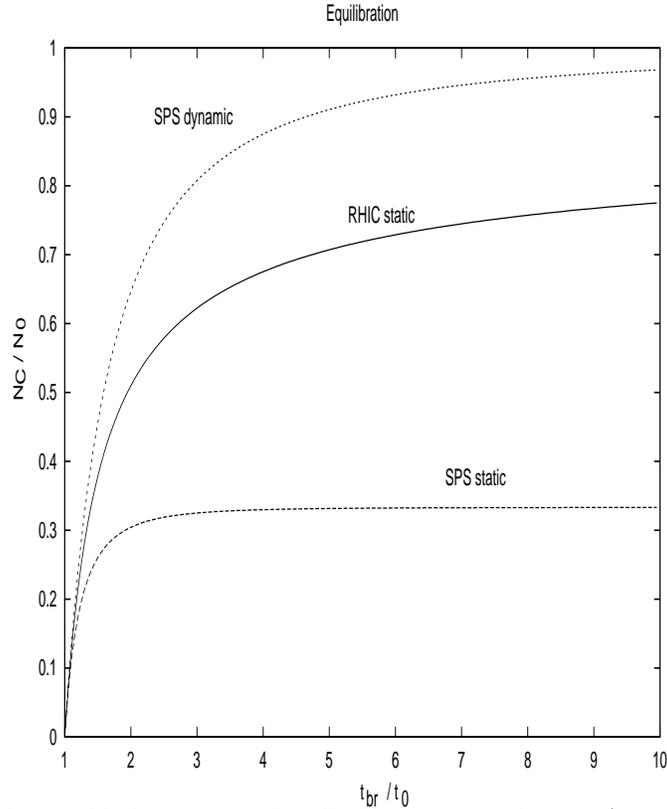,width=110mm,height=90mm}} }


\label{Fig1}
\caption{
Degree of the chemical equilibrium reached for different
break-up times $t_{{\rm br}}/t_0$ and initial inter-collison times $t_c/t_0$
for SPS  and RHIC expansion scenarios with static and dynamic confinement.
}
\end{figure}


\vs
Finally we compare characteristic values for SPS
and RHIC. 
In the SPS scenarios we assume  $t_0=1$ fm/c as the starting 
time of the expansion. The collision time changes somewhat due to
different temperatures (and effective quark masses) presumed at
the beginning of hadronization. 
We consider a value characteristic for quark-gluon plasma,
$t_c \approx 0.33$ fm/c.
We take over the ratio $t_0/t_c=3$ for the RHIC scenario, too.
The break-up time is the most unknown factor, therefore
the equilibration has been plotted as function of $t_{{\rm br}}/t_0$.
Fortunately, once the stauration value of $\gamma_C$ has been approached,
there is no strong dependence on it.
The $1-e^{-1} \approx 0.63$ fraction of equilibrium value is reached
after $2$ fm/c for the dynamic SPS, after $3.5$ fm/c for the
static RHIC scenario. In the static SPS scenario it will be never
reached, $\gamma_C$ saturates near $0.3$ already after $2$ fm/c.
This last scenario, quark coalescence with density independent
probability, can be excluded on the basis of quark confinement.


\vs

\vs
In conclusion
1) the results of the statistical model can be interpreted not only by 
assuming hadronic equilibrium, but as well by quark -- hadron mixture
equilibrium with quarks eventually suppressed due to confinement. 
This interpretation
avoids the collison time problem: the hadrons need not transmutate into
each other directly.
2) The ALCOR and constrained hadronic equilibrium leads to the 
same result in this
simple scenario due to too many constraints on conserved quantities.
Quark coalescence, which takes conservation into account, therefore mimics
hadrons in chemical equilibrium.
This feature certainly does not not generalize for several competing reaction
channels, but the overall picture can be similar. ALCOR actually
assumes more conserved quantities, than usual: quark and antiquark 
numbers in any flavor are conserved separately due to the assumption
of rapid hadronization.
3) $\gamma_C \approx 0.3$ is typical for CERN SPS without
considering dynamical confinement,
even for an arbitrary long hadronization process.
This leads to a factor two to three characteristic
error on equilibrium model results.
The experimental evidence, reaching at least 70-80\% of equilibrium
ratios for almost all hadrons in a short time in the order of
several fm/c, excludes this possibility. It also underlines
the quark matter hadronization scenario with dynamical confinement
effect, as transchemistry describes it. 
4) An expansion scenario typical for RHIC and LHC energies in principle
allows for reaching the equilibrium even with static confinement, but 
the process is rather slow and an early break-up may
prevent from hadronizing all quarks. Dynamical confinement has to be
in work for one-dimensional expansion scenarios, too.

\vs
One is tempted to speculate about the origin of dynamical confinement,
looking for a cause of the density dependence of the hadronization
cross section. Since this dependence makes the cross section stronger
with diminishing quark density, it cannot have perturbative origin;
on the contrary it has to do with formation of strings or color ropes.
As the color density drops, strings, follow-ups the chromoelectric flux
lines, become longer before ending on a partner color. 
The volume, in which possible partner to form a hadron occur, hence
increases, so does the hadronization cross section,  $\sigma \sim \ell^3.$
The characteristic string length on the other hand depends on the
color density as $\sim n^{-1/3}_{{\rm color}}$. Such a mechanism
would lead to $\sigma = \sigma_0 n_{{\rm color}}(0) / n_{{\rm color}}$.

\vs
{\bf Acknowledgements}
Discussions with J.~Zim\'anyi, K.~Redlich and U.~Heinz
are gratefully acknowledged.
This work was supported by the US-Hungarian Joint Fund
T\'eT 649, by the Deutsche Forschungsgemeinschaft DFG-MTA 101 
and by the Hungarian National Research Fund OTKA T029158.


\vs


\end{document}